# Electrical properties of ferroelectric $YMnO_3$ films deposited on n-type Si (111) substrates

S Parashar, A R Raju and C N R Rao*

Chemistry and Physics of Materials Unit, Jawaharlal Nehru Center for Advanced Scientific Research, Jakkur, Bangalore 560 064, India

P Victor and S B Krupanidhi

Materials Research Center, Indian Institute of Science, Bangalore 560 012, India

## Abstract

$YMnO_3$ thin films were grown on n – type Si substrate by nebulized spray pyrolysis in Metal – Ferroelectric – Semiconductor (MFS) configuration. The C-V characteristics of the film in MFS structure exhibit hysteretic behavior consistent with the polarization charge switching direction, with the memory window decreasing with increase in temperature. The density of interface states decreases with the increase in the annealing temperature. Mapping of the silicon energy band gap with the interface states has been carried out. The leakage current measured in the accumulation region, is lower in well-crystallized thin films and obeys a space- charge limited conduction mechanism. The calculated activation energy from the dc leakage current characteristics of Arhennius plot reveals that the activation energy correspond to the oxygen vacancy motion

* author of correspondence; electronic mail: cnrrao@jncasr.ac.in Fax # +91-80-8462766

## 1. Introduction

Rare – earth manganites of the type $Ln_{1-x}A_xMnO_3$, where Ln is a rare earth and A is an alkaline earth, have become known for a variety of fascinating properties, such as colossal magnetoresistance, charge ordering and orbital ordering [1-2]. These properties are strongly influenced by the average size of the A-site cation, $\langle r_A \rangle$ [3]. The parent manganites of $LnMnO_3$ also show marked effects of the size of A-site cation. $LaMnO_3$ is an antiferromagnetic insulator where the Jahn-Teller distortion around $Mn^{3+}$ plays an important role. However, as the $\langle r_A \rangle$ becomes small, the perovskite structure becomes unstable at ambient conditions, resulting in a hexagonal structure. Thus, $YMnO_3$ crystallizes in the $P6_3cm$ space group [4] with, a = 6.130Å and c = 11.505 Å. The structure is best described as layers of manganase-centred trigonal bipyramids with 5-coordinate Mn but no framework of Mn-O bonds along the axial direction (c-axis). The equatorial oxygens are corner-shared by three polyhedra in the basal plane. This structure leads to an unipolarization axis along [0001], which can be reversed by an electric field, thereby making these manganites exhibit ferroelectric behavior. Although ferroelectrcity as such in the bulk form was discovered in 1963 by Bertaut et al [5], there has been only recent effort to prepare them in thin – film form [6] to investigate ferroelectricity. Various techniques like MBE (moleculer beam epitaxy), PLD (pulsed laser deposition) and sol-gel method have been employed to deposit these films. We considered that it would be most worthwhile to explore a simpler chemical route to provide alternative or even better means of depositing films of novel complex oxide. With this in mind, we have employed nebulized spray pyrolysis of organometallic precursors to deposit films of $YMnO_3$. In addition, it is also noteworthy that $YMnO_3$ is also antiferromagnetic, thus making it a biferroic [7].

Interest in $YMnO_3$ and related compounds, arises from their potential use in ferroelectric random access memories (FRAMs) in the metal-ferroelectric-semiconductor field-effect transistor (MFSFET) mode [8]. An additional advantage of these manganites is the absence of volatile elements such as Pb and Bi. This is especially useful in MFSFET applications as one can directly integrate onto Si substrates. In the present study, we have deposited films of $YMnO_3$ on Si (111) substrates and conducted a detailed study of their ferroelectric properties in terms of C-V characteristics.

**2. Experimental**

Thin films of $YMnO_3$ were prepared by nebulized spray pyrolysis [9]. The technique involves the pyrolysis of a nebulized spray of organometallic derivatives of the relevant metals. The atomized spray is slowly deposited on a solid substrate at an appropriate temperature, with sufficient control of the rate of deposition. Nebulized spray pyrolysis is a gentle method since the decomposition of the atomized spray to yield the oxide occurs at sufficiently low temperatures to be favorable to maintain the desired stoichiometry. We have carried out nebulized spray pyrolysis by employing a simple locally fabricated apparatus. A pizeo-electric transducer of 20mm diameter operating at 1.72 MHz was used to carry out the deposition. The solution is in direct contact with the liquid to be sprayed. We have employed the acetate of Y and the acetylacetonate of Mn as precursors in the present study. Solutions of stiochiometric quantities of the precursors were prepared in methanol solvent. $YMnO_3$ films of ~1µm to 2µm thickness were deposited on the n-type Si(111) substrates (~$\rho$ = 1 $\Omega$ cm, dopant concentration 6.5 x $10^{14}$ $cm^{-3}$) at 450 K by using air as the carrier gas (flow rate 1.5 liters/min). The films so obtained were annealed in air at 973 K (denoted as Y33), 1073 K (denoted as Y73), and 1123 K (denoted as Y123) for 6 hrs each in order to study the process of crystallization as a function of annealing temperature. The thin films were characterized by X-ray diffraction (XRD), energy dispersive analysis of X-rays (EDAX) and scanning electron microscopy (SEM). The electrical measurements were carried out by sputtering gold electrodes on the $YMnO_3$ thin films as top electrodes for the Metal - Ferroelectric – Semiconductor (MFS) configuration. The dielectric and capacitance – voltage (C – V) measurements were carried by using HP 4284 A LCZ meter to measure capacitance in the range varying from 100 Hz to 1 MHz at oscillating voltage of 500 mV. The current - voltage measurements were measured using Keithley SMU 236 at elevated temperatures ranging from 300 K to 473 K.

## 3. Results and Discussion

### *3.1 Crystallinity and Composition analysis*

Fig.1 shows the X – ray diffraction (XRD) patterns of the $YMnO_3$/Si (111) films annealed at different temperatures. The XRD patterns reveal that the crystallinity of the films increases as the annealing temperature is increased from 973 K to 1123 K. The evolution of the polarization axis is also seen from the increase in the intensity of the (0004) reflection. Fig.2 shows the cross sectional SEM of the $YMnO_3$ thin films crystallized at 1123 K and its morphology is shown in the inset. The morphology was found to be dense and the interface was found to be uniform and coherent from the cross – sectional images. A semi-quantitative analysis of the composition of $YMnO_3$ thin films was found using energy dispersive analysis of X – rays (EDAX) and the thin films maintained an excellent stoichiometric ratio.

### *3.2 Electrical Characteristics*

Dielectric studies and dc leakage current characteristics of the $YMnO_3$ films deposited on n–type Si substrate were carried out in the Metal – Ferroelectric – Semiconductor (MFS) configuration.

#### *3.2.1. C – V characteristics*

C – V measurements were done from 100 Hz to 1 MHz for all the well crystallized polycrystalline $YMnO_3$ thin films at elevated temperatures ranging between 300 K to 500 K. The dielectric constant was calculated on taking the account of the thickness of native oxide formation on the silicon based on the annealing conditions and dielectric constant was calculated for 1 MHz in the accumulation region. The capacitance – voltage measurements, the oxide capacitance was determined by using the nonlinear least-square fit algorithm developed by Hauser and Ahmed [11]. Quantum mechanical corrections were not applied, as the films were sufficiently thick, consistent with the work of McNutt and Sah [12]. The dielectric constant obtained for the $YMnO_3$ thin films on Si in the accumulation region was 25, a value close to that reported by Yoshimura et.al [6]. The dielectric constant was found to increase with the annealing temperature. This enhancement of the dielectric constant in the annealed films may be associated with the improved crystallinity, lesser porosity and lesser defects of the thin films.

The gate voltage was swept from $-V_g$ to $+V_g$ and vice versa resulted in a hysteresis behavior. Counter clockwise hysteresis was observed due to the ferroelectric polarization-switching behavior as shown in fig.3 for Y123 films [10]. As the bias sweep rate was decreased, the present MFS structure exhibited the same magnitude of hysteresis window, thus demonstrating good memory retention characteristics of the $YMnO_3$ films on n–type Si, unlike the memory window collapse in case of p–type Si reported by Yoshimura et.al [6].

When a ferroelectric film is integrated onto a Si – substrate, the polarization charge density in contact with the Si surface, modifies the surface conductivity. The magnitude of the modulation depends on the applied electric field and follows the normal hysteresis behavior of the ferroelectric. K Ito and H Tsuchiya clearly identified reversal modes of the memories of ferroelectric thin films integrated on a semiconductor. Polarization mode memory operation clearly establishes the ferroelectric nature of the films on silicon. This approach was taken in the present studies to establish the presence of ferroelectricity in $YMnO_3$ thin films. Polarization mode memory behavior was studied in terms of C-V hysteresis at different temperatures and results are shown in the inset of fig.3. The polarization charge density is expected to reduce as one approaches the Curie temperature, resulting in narrowing or the disappearance of the memory window. This behavior is clearly observed in the present studies, thereby establishing ferroelectricity in the $YMnO_3$ films.

The C – V curves measured at 1kHz and higher were of high frequency type with minimum capacitance at the inversion region. The minority carriers constituting the inversion layer do not respond for frequencies higher than 1 kHz, as the response time of the inversion layer is very long (of 0.01 – 1sec for p-type Si ), compared to the time scale of applied ac signal at the room temperature. This indicates that the ac signal is sufficient to induce generation and recombination of the minority carriers in the depletion layer of the bulk n–type Si, leading to charge exchange with the inversion layer. In our case, the inversion layer was able to respond at the lower frequencies (~ 100 Hz) in Y123 thin films as shown in the fig.4. The flat band capacitance was calculated using the following expression

$$C_{fb} = \left( \frac{1}{C_{ox}} + \frac{1}{C_{Ld}} \right)^{-1} \qquad (1)$$

with $C_{ox}$ denoting the maximum oxide capacitance in the accumulation region,

$$C_{Ld} = \left( \frac{\varepsilon_s A}{L_d} \right)$$

$\varepsilon_s$ denoting the dielectric constant of the silicon, A denoting the electrode area and $L_d = \left( \frac{\varepsilon_s kT}{q^2 N_a} \right)^{1/2}$. The oxide charges and interface states in a MOS capacitor, plays a crucial role for an efficient performance and to exhibit a thermally stable device. The $V_{FB}$ is calculated as follows:

$$V_{FB} = \phi_{MS} - \frac{Q_I}{C_{OX}} \qquad (1)$$

with $\phi_{MS}$ denoting the work function difference between metal and semiconductor, $C_{OX}$ as the oxide capacitance and $Q_I$ the effective interface charge [14]. The flat band voltage ($V_{FB}$) [13] was found to shift towards zero voltage with increasing annealing temperature due to better crystallinity as shown in the Table 1. Variation of $V_{FB}$ in the annealed thin films is due to the structural changes in the oxide - semiconductor interface and it reflects the changes in the density of defect states in the insulator. The $V_{FB}$ is found to decrease with increasing oxidation temperatures, indicating a reduction and/or compensation of fixed charges.

There are four general types of charges associated in $SiO_2$ - Si system, such as, fixed oxide charges $Q_f$, mobile oxide charges $Q_m$, oxide trapped charges $Q_{ot}$ and interface trapped charges $Q_{it}$. It may be realized that the $Q_f$ and $Q_{it}$ are vital in degradation of the device. Fixed oxide charges are positive, stable in the oxide and lie close to $SiO_2$ - Si interface that cannot be charged or discharged by varying the silicon surface potential. Interface states are located at $SiO_2$ - Si interface and are in direct electrical communication with silicon, and responds rapidly to the changes in the silicon surface potential. Hence the interface states were calculated from the C –V analysis. In order to minimize the $D_{it}$, an annealing was carried out at higher temperatures in the ambient of dry oxygen. The low temperature annealing was avoided because it induces a transfer of hydrogen, which eventually activates and annihilates the interface traps. [17]

The interface states arise due to the dangling bonds or unsaturated bonds at the interface of Si and $YMnO_3$ thin films. The stretch out of the C–V curve at higher frequencies along the voltage axis for the $YMnO_3$/n-Si film is not appreciable, indicating the minor amounts of interface traps present. The interface trap density was obtained by the Castagne–Vaipalle method [15] for the films annealed at 1123 K, as they exhibited both low and high frequency curves as shown in the fig.3. The density of interface states was calculated from the C–V curve as follows:

$$N_{SS}(\psi_S) = \frac{1}{qA}\left(\frac{C_{OX}C_{LF}}{C_{OX} - C_{LF}} - \frac{C_{OX}C_{HF}}{C_{OX} - C_{HF}}\right) \qquad (2)$$

Here $\psi_s$ is the surface potential, $C_{ox}$ is the oxide capacitance, $C_{LF}$ is capacitance at low frequency (100 Hz), $C_{HF}$ is capacitance at high frequency (1MHz), q is charge of electron and A is the area of electrodes. The densities of the interface traps for films annealed at 1073 K and 973 K was calculated using high frequency Terman method [16]. The calculated values of $N_{SS}$ are shown in Table 1. The mapping of the energy band gap of Si with the density of interface states of $YMnO_3$ thin films crystallized at 1123 K is shown as an inset of the fig.4. The density of the interface states is lower than that reported earlier [6] while the $N_{ss}$ is higher at the band edges with a dip at the center of the energy band gap of Si for $YMnO_3$ films annealed at 1123 K.

### *3.2.2. DC leakage current characteristics*

Various transport mechanisms of charge flow are involved in the metal – ferroelectric – semiconductor (MFS) configuration. The mechanisms include Poole – Frenkel [17] and Schottky effects [18] resulting from the lowering of a columbic potential barrier by the applied electric field in bulk and electrode – bulk interface respectively, and field ionization of trapping levels known as space charge limited current (SCLC) [19, 20]. The dominant current conduction mechanism was analysed from the dc leakage current behavior. Well-crystallized films of $YMnO_3$ (annealed at 1123 K) exhibited lower leakage current than those annealed at lower temperatures as shown in fig.5. These $YMnO_3$ thin films didn't obey the Schottky and Poole Frenkel effects as the calculated high frequency dielectric constant from their respective conduction equations were of one order of magnitude less than the dielectric constant of the $YMnO_3$ bulk ($\varepsilon_\infty$ ~ 20).

### *(i) Space Charge Limited Current mechanism*

A space charge region will be set up in the conduction band on an inexhaustible supply of free carriers in the dielectric near the injecting electrode on assuming the ohmic contact has been formed at the injecting electrode. The space charge limited current for an insulator without traps, obeys Child's law given as

$$J = \frac{9}{8}\frac{\varepsilon\mu}{w^3}V^2 \qquad (9)$$

where ε is the dielectric permittivity, μ the drift mobility of electrons , w the thickness of insulator layer and V the voltage drop across the layer. According to the SCLC model, the J is proportional to $V^2$ and from the fig.5, it shows that J is proportional to $V^2$ over a significant range of voltage in the region III. The slope of J(V) in the non - ohmic region is not the same at different temperatures. Suppose electron traps influence the SCLC, then the current obtained is trap filled limited current and these current arises due to both shallow and deep electron traps. $YMnO_3$ films annealed at 1123 K obey the space-charge conduction limited mechanism (SCLC) as shown in the fig.5. The J – E plot shows a near linear dependence (region I) upto voltage trap filled limited ($V_{TFL}$ ~ 3.4 V) and curves spikes up with a slope of 15 –18 (region II) and again saturating to a slope of 2 (region III). The sudden increase of current in the *I –V* plot was obtained repetitively, which ruled out the possibility of breakdown phenomenon to be the reason for the sudden increase of current above $V_{TFL}$. In the region II (α ~ 16.1) there is a leap in the slope at voltage trap filled limited ($V_{TFL}$) which is due to the filling up of the trap levels present in the $YMnO_3$ thin films and in the region III (α ~ 2) it exhibits trap free square law

$$J = \left( \frac{9k\mu\varepsilon_o E^2}{8L} \right) \qquad (3)$$

with the J is the current density, k is the Boltzmann constant, μ is free electron mobility, $\varepsilon_o$ is the permittivity of vacuum, E is the applied electric field and L is the thickness of the film.

***(ii) Trap filled voltage ($V_{TFL}$)***

It is known in space-charge theory that trap filled voltage is a voltage where the Fermi level would increase well above all deep trap levels. Above the trap filled limit, all those trap levels would remain filled at a given temperature. If the temperature was increased, few of the filled traps would re-emit some electrons from the trap sites, and again those sites would become empty. In other words, the ratio of the free electrons to the trapped electrons would increase with temperature. As a result, one would have to apply a higher voltage to inject further electrons in the sample, so that all the trap levels get filled with electrons. This would be the possible cause for the increment of $V_{TFL}$ with temperature. But this is not true in our case as $V_{TFL}$ decreases with the increase in the temperature from 300 K to 450 K as shown in the fig.6 (for Y123 films). This remains opposed to the space charge theory given by Lampert as it deals on the conduction mainly by electronic charges. It was assumed that the space charges would be represented by an equilibrium distribution which would be identical to the Fermi–Dirac distribution function, at all temperatures. To achieve equilibrium, a certain time is required. The charges, immediately after being injected into the sample, have to equilibrate with the surrounding, to reach the time independent distribution of electrons among various energy levels. This has to occur throughout the entire sample.

For that, the electrons have to migrate uniformly through the sample, so that they could seek the equilibrium (trap) sites. This means that the space-charge transient would govern the phenomenon of the trap distribution if observed in a short time scale. It is known that the space-charge transient in thin films could be of a time scale of several seconds depending upon the sample. The electron distribution would therefore be limited by the competition between the rate of trapping and detrapping of the electrons. The trapping rate would be represented by

$$T_{E_c \to E_t} = N(E_c)\sigma(E_t)N(E_t)$$

where, ‘‘$T$’’ represents the rate of transition between the energy states mentioned in the suffices, ‘‘$N$’’ represents the density of unoccupied energy states at energy ‘‘$E$’’, and ‘‘$\sigma(E_t)$’’ is the capture cross section of the traps. The detrapping of electrons would be given by

$$T_{E_c \to E_t} = \nu N(E_t)\exp(\Delta E / kT)N(E_c)$$

where the energy difference has been denoted by ‘‘$\Delta E$’’ and the attempt frequency to escape is ‘‘ $\nu$’’. At a lower temperature, there would be very few electrons in the upper energy state (from where, it is captured to the trap sites), and the trapping rate also would not change much with temperature. Therefore, the entire process would be limited by the detrapping only. But at a higher temperature, there would be a significant amount of electrons in the conduction band (due to thermal generations and so on) which in turn would increase the rate of trapping, causing a greater number of trapped electrons.

There would be some contribution from the detrapping of electrons also, but the increased trapping rate might overshadow the detrapping rate at a higher temperature. Therefore, one would expect that, the number of electrons required to fill all the traps would be less at a higher temperature, than it was at a lower temperature. This might bring down the trap filled voltage closer to the actual value. This should saturate at the equilibrium value of $V_{TFL}$. There are reports of the space-charge transients in insulating samples even in this time scale. This was attributed to the movement of oxygen vacancies, which are quite slower compared to electrons. The existence of the oxygen vacancies was already established in the preceding section. It could, therefore, be assumed that both electrons and oxygen vacancies dominated the space-charge phenomenon. At lower temperatures, oxygen vacancies somehow did not respond to higher fields (the nonlinear space-charge region), and the highly mobile electrons were the majority charge carriers of the current. Due to their higher mobilities, they almost instantaneously redistributed themselves within the sample according to the space-charge law, and therefore, the expected increase of $V_{TFL}$ with temperature was noticed. However, at high temperatures, even the oxygen vacancies played a significant role in the space-charge conduction. [21] Since, oxygen vacancy motion is a slower process, it required significant time to respond, and this might be the reason why the linear to nonlinear region transition voltage (or the trap filled voltage) decreased with temperature, or with the increase of delay time. The current in the linear region was found to fit with the Arrhenius equation:

$$I \sim \exp(-\Delta E / kT)$$

with ΔE denoting the activation energy obtained, and the calculated activation energy was of 0.92 eV for the 1.1 V on the Y123 thin films as shown in the fig.7. This order of magnitude of the activation energy can be due to the motion of the oxygen vacancies as reported by many authors. The calculated trap charge concentration was found to decrease with an increase in the temperature as shown in the fig.8 (for Y123 films).

**4. Conclusions**

$YMnO_3$ films have been deposited successfully on n – type Si by nebulized spray pyrolysis. The bulk dielectric constant was found to be 25 from the accumulation capacitance. The C-V hysteresis behavior and the rotation direction established ferroelctricity in these $YMnO_3$ films. The reduction in the width of memory window with ambient temperature further confirms the ferroelectric behavior of the $YMnO_3$ films. The interface states have been studied and the densities of states were estimated. A detailed analysis on the dc leakage current analysis reveals to be space charge conduction limited (SCLC) process and the obtained activation energy was attributed to the oxygen vacancy motion.

**5. Acknowledgement**

The authors thank BRNS (DAE), India, for support of this research. One of the authors (P.V) wishes to acknowledge CSIR, India for the senior research fellowship.

**Table I**

Flat band voltages and densities of interface traps in $YMnO_3$ thin films.

| Annealing temperature ($^oK$) | $V_{FB}$ (V) | $N_{ss}$ ($eV^{-1}$ $cm^{-1}$) |
|---|---|---|
| 973 | 0.505 | $3.2 \times 10^{12}$ |
| 1073 | 0.2275 | $9.1 \times 10^{11}$ |
| 1123 | 0.198 | $5.1 \times 10^{11}$ |

**Figures**

Fig.1 X–ray diffraction pattern of the $YMnO_3$ thin films on Si(111) annealed at (a) 973 K (b) 1073 K (c) 1123 K.

Fig. 2. Cross-sectional SEM micrograph shows that the interface is uniform and coherent for a Y123 film. Inset shows the morphology of the same film.

Fig. 3 C-V characteristics of $YMnO_3$ (Y123) film on n-type Si(111) at different bias sweep rates: 0.75 $Vs^{-1}$(squares), 0.375 $Vs^{-1}$ (circles), 0.093 $Vs^{-1}$ (up triangles). The inset shows the C-V characteristics of the films at different temperatures. The loop are a decreases as the temperature is increased: 297 K (circles), 313 K (down triangles) and 373 K (squares).

Fig. 4 C–V characteristics of $YMnO_3$ (Y123) thin films on n – type Si(111) measured at 100 Hz (up triangles) and 1 MHz(circles) at 300 K. The inset shows the mapping of the density of interface states with the energy band gap of Si.

Fig. 5 Log I – Log V curves in the accumulation region in MFS structure for the different annealing temperature, .i.e. 973 K (squares), 1073 K (up triangles) and 1123 K (circles).

Fig. 6. Voltage trap filled limited ($V_{TFL}$) as function of temperature of the Y123 thin films

Fig. 7. Arhennius plot of current versus temperature at different voltages of Y123 thin films.

Fig. 8. Trap charge concentration versus temperature of Y123 thin films.

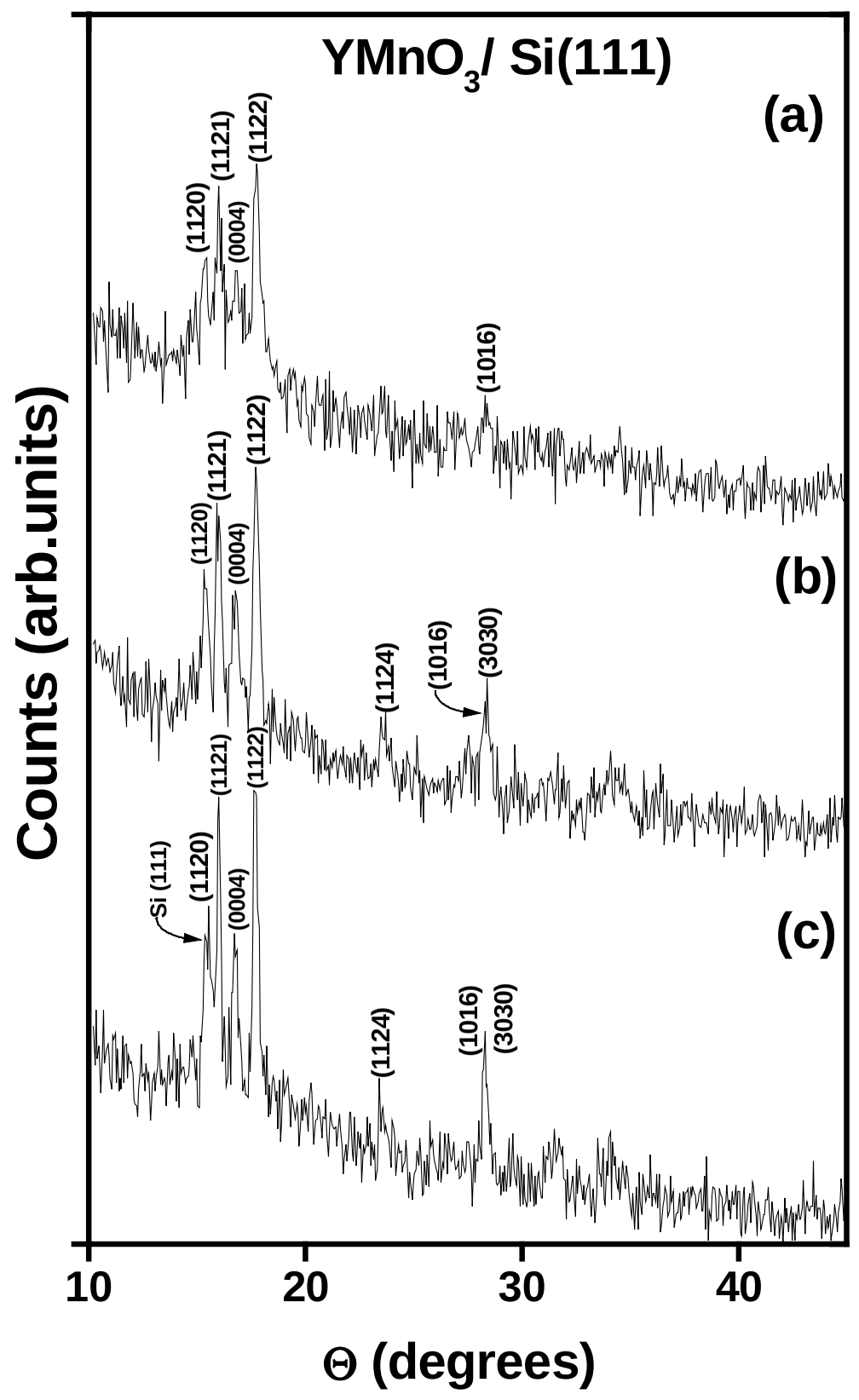

**Fig. 1**

**Parashar *et al.***

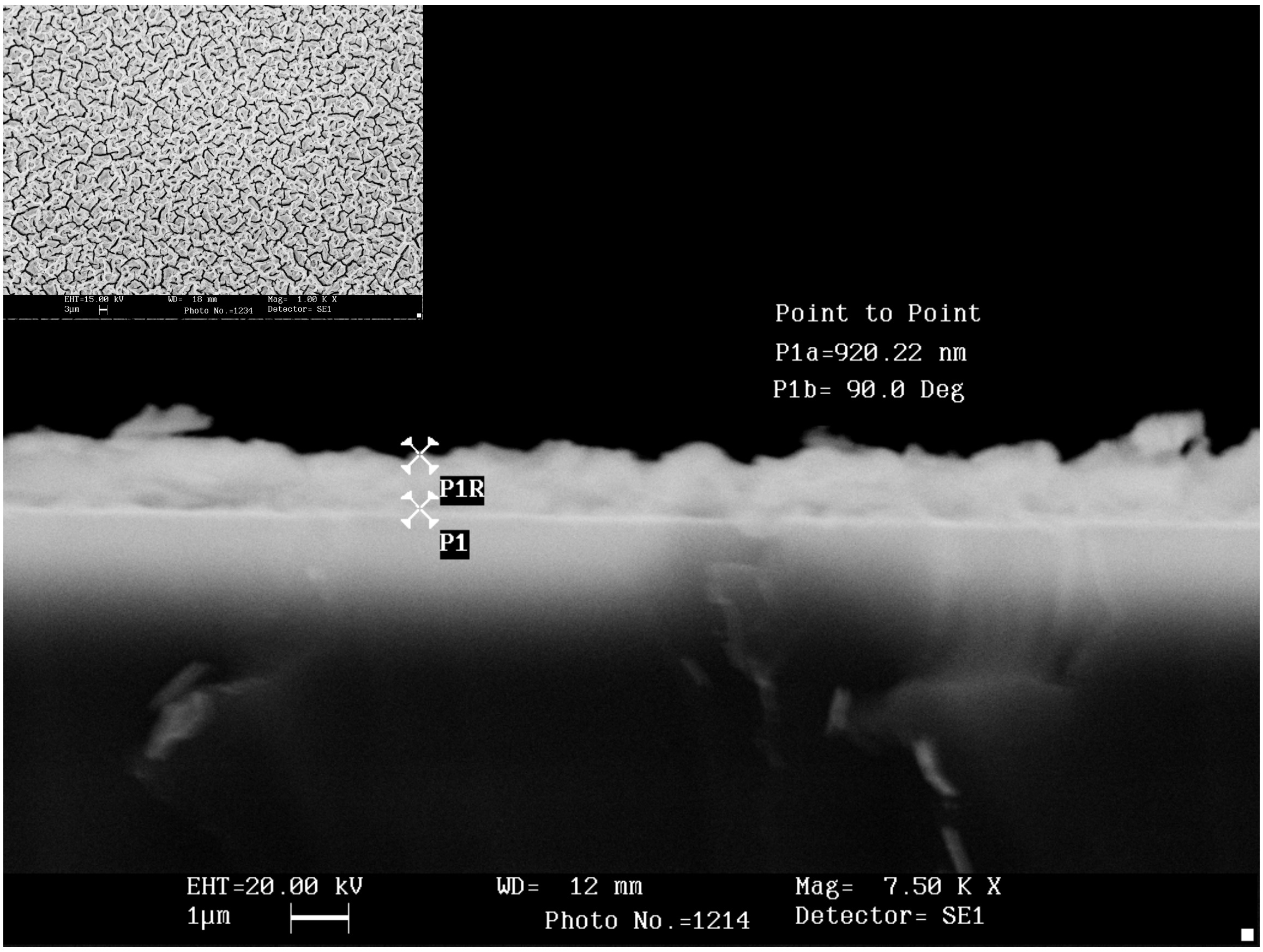

# Fig.2 Parashar et al

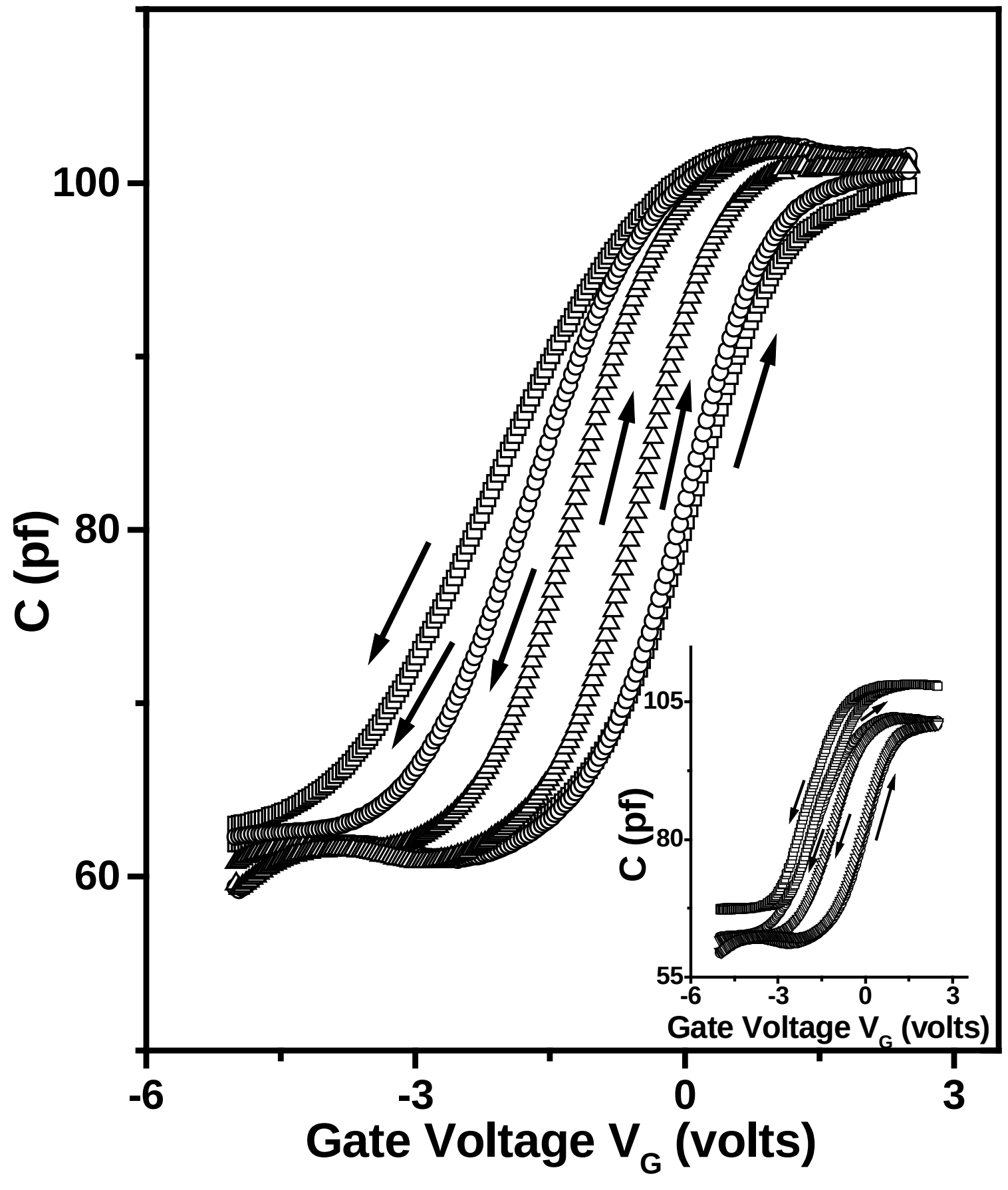

**Fig. 3**
**Parashar *et al.***

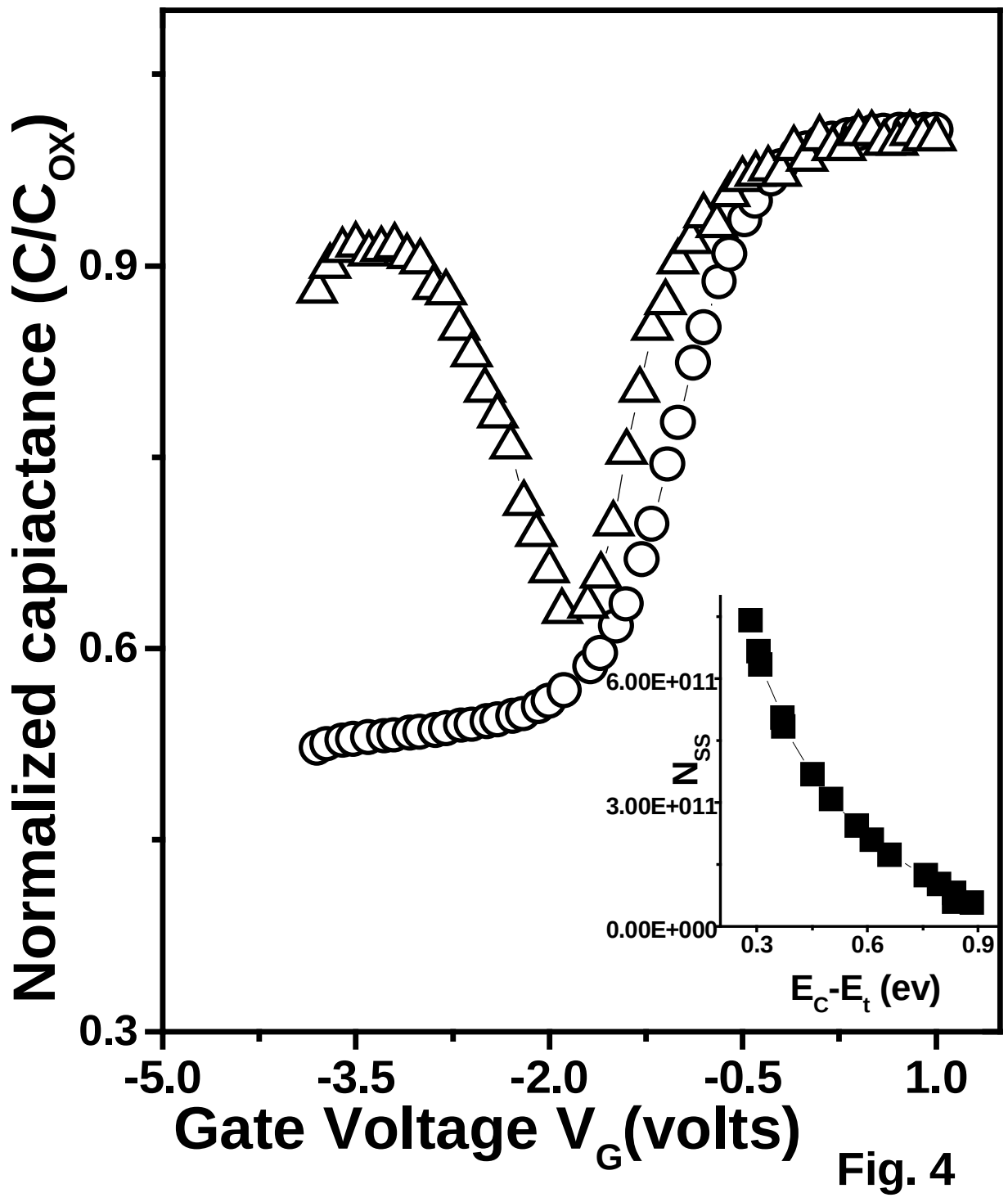

**Fig. 4**
Parashar *et al.*

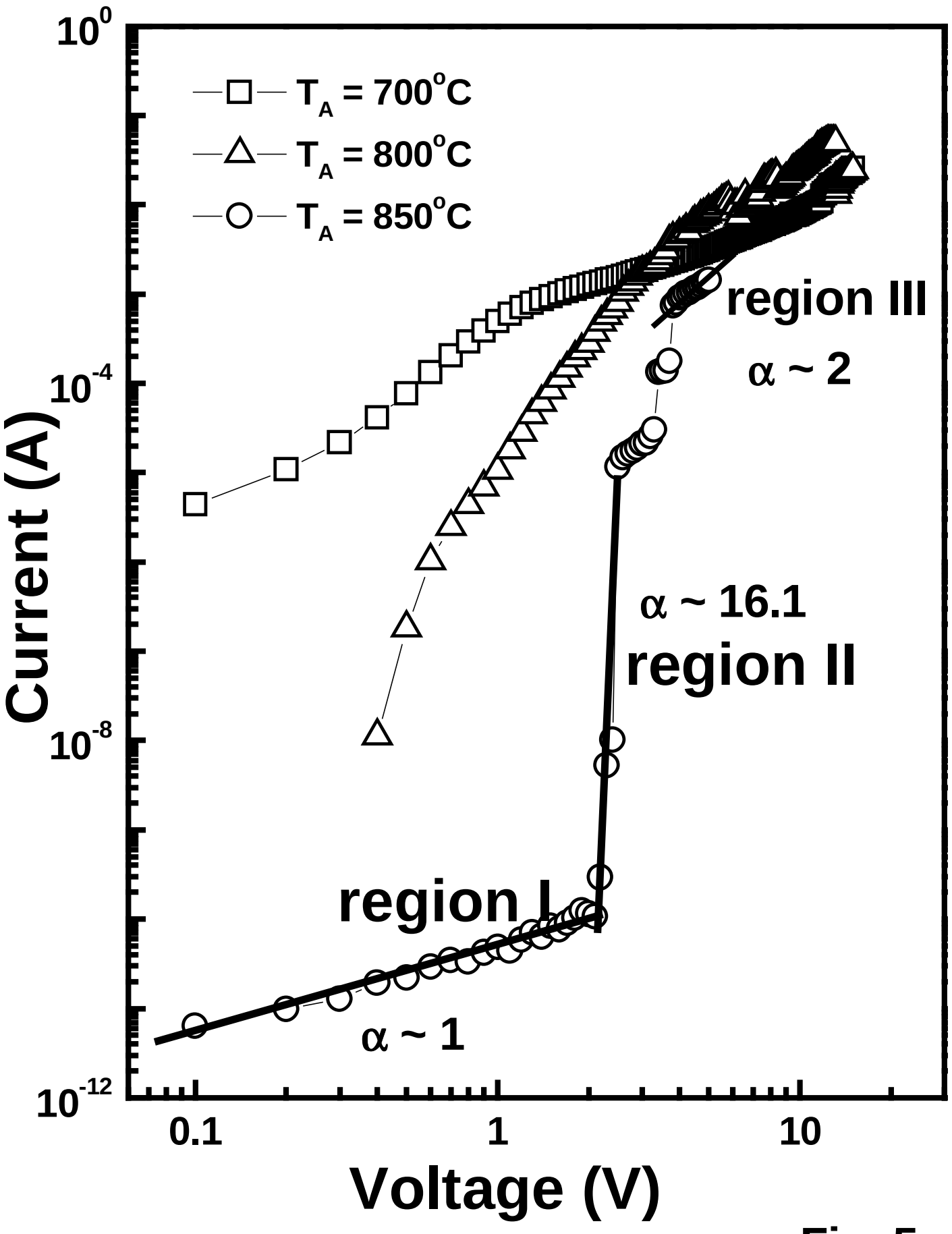

**Fig. 5**

**Parashar *et al.***

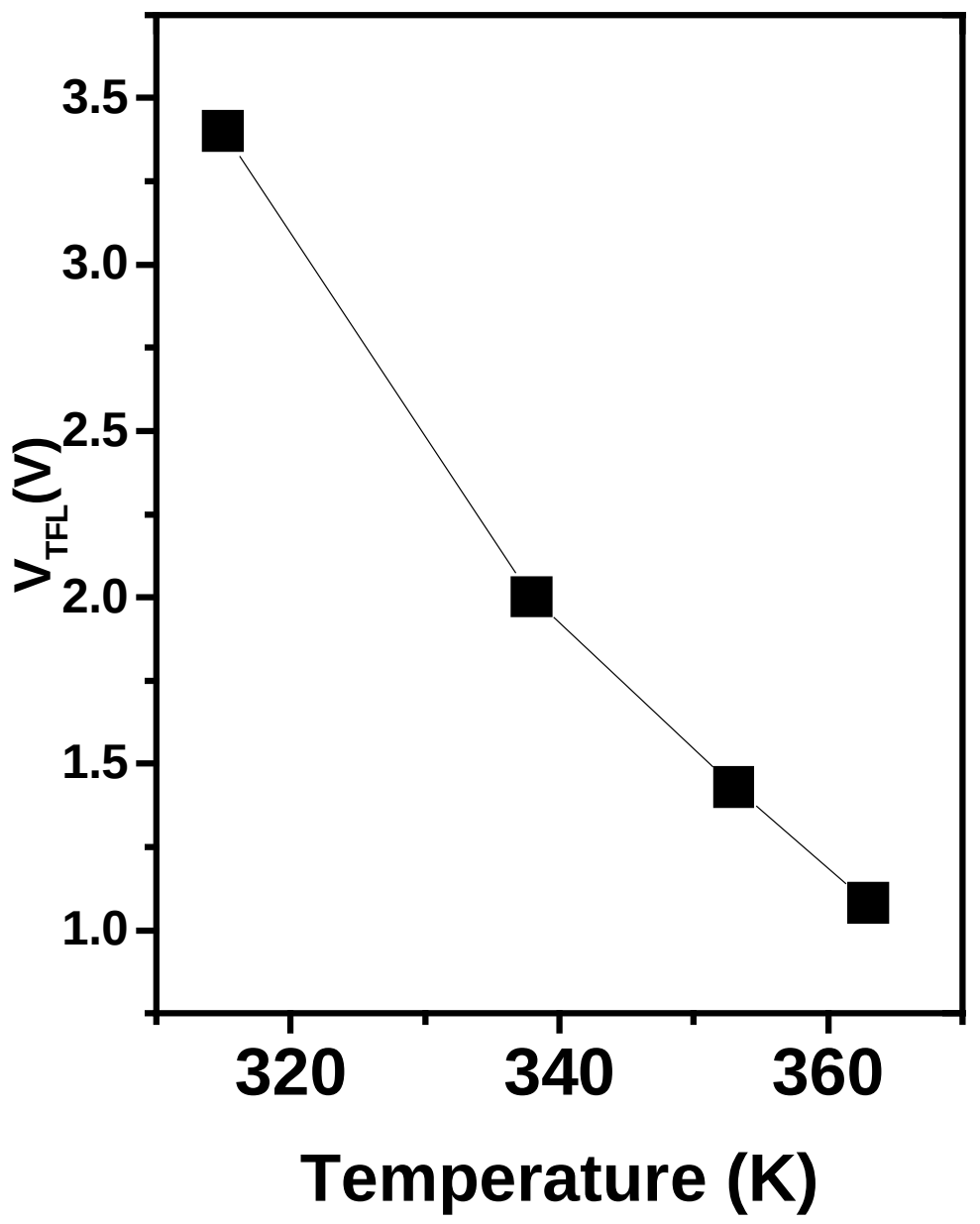

**Fig. 6**
**Parashar *et al.***

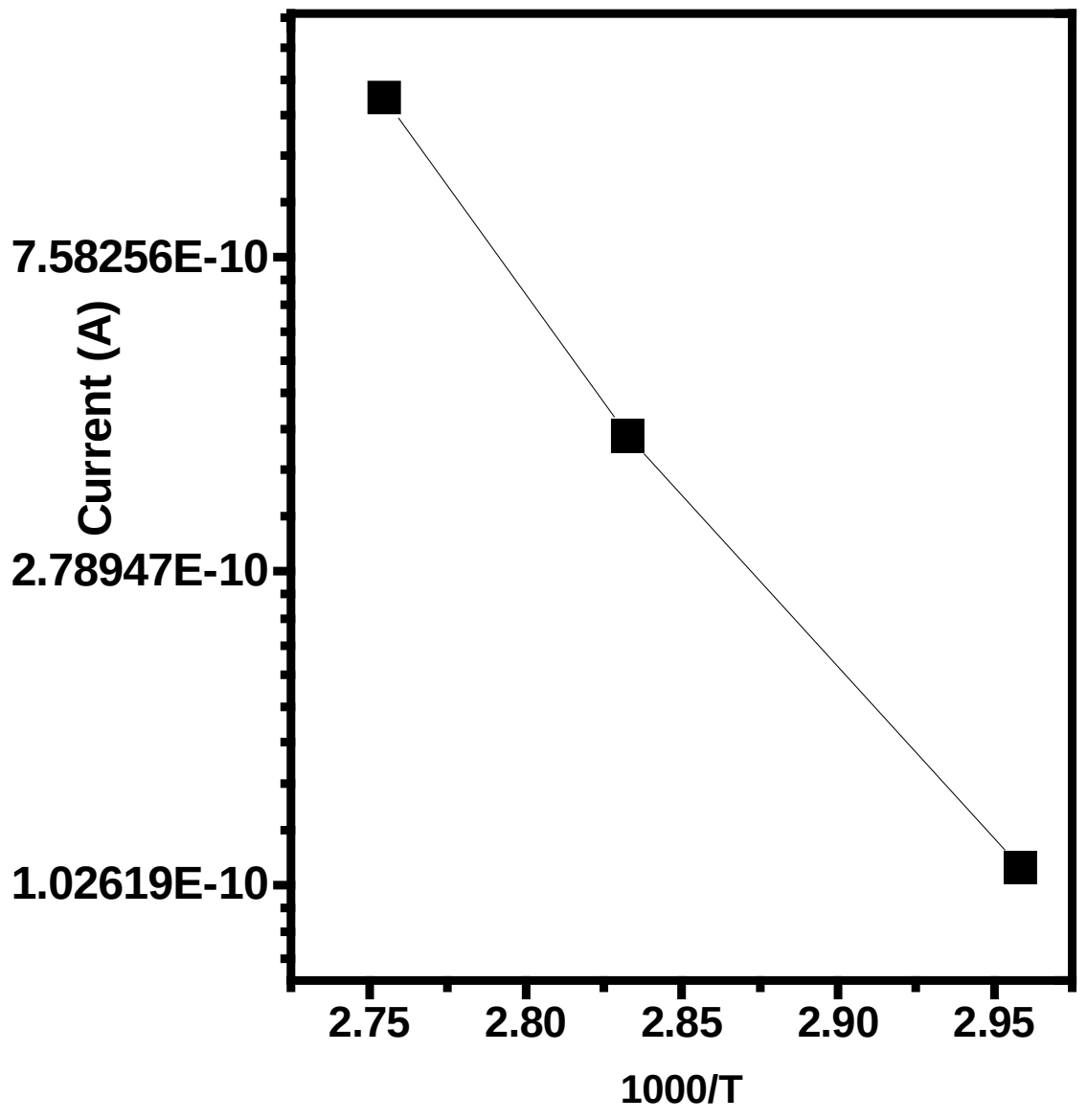

**Fig. 7**
**Parashar *et al.***

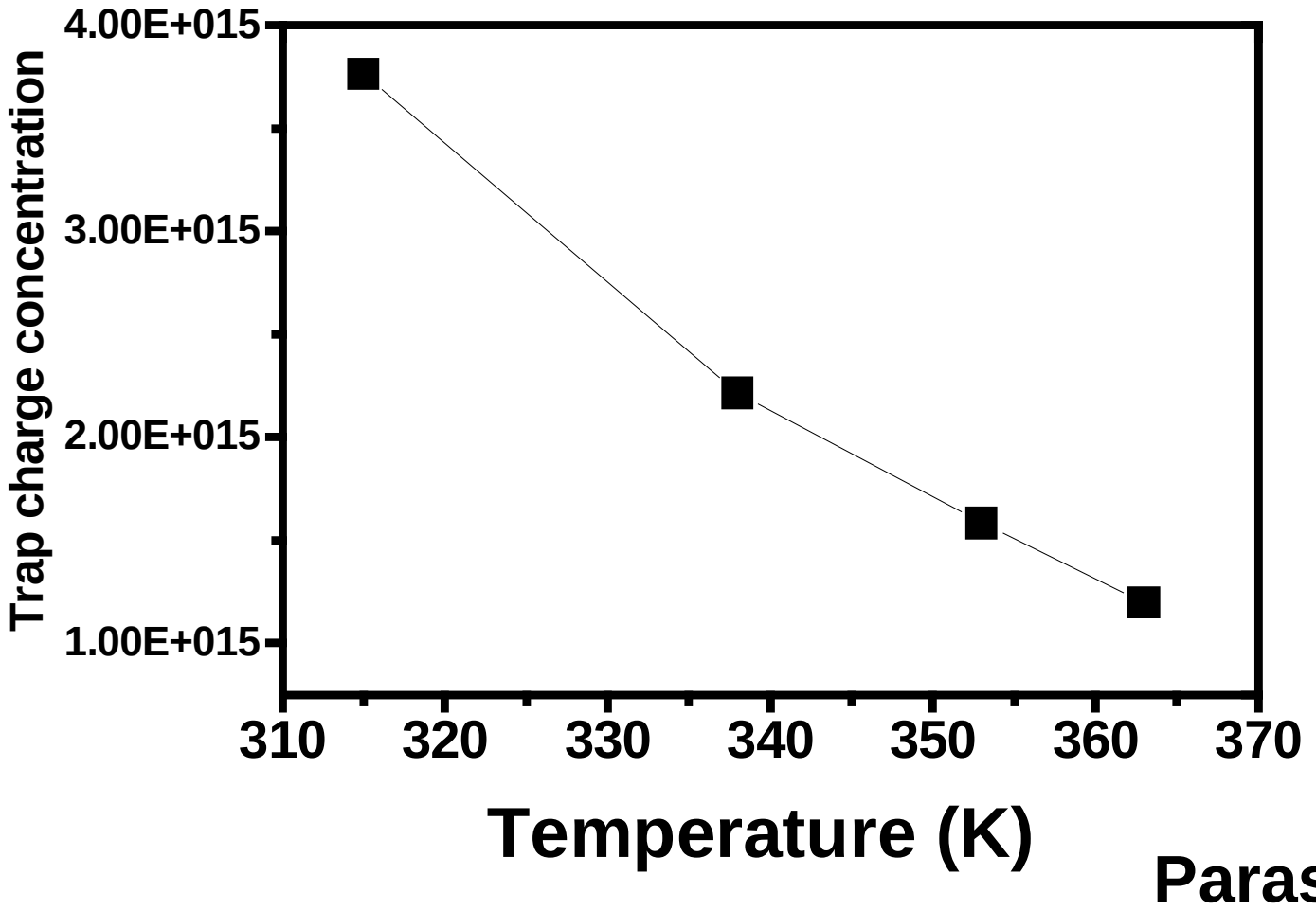

**Fig.8**
**Parashar *et al.***